# High return loss at the end face of fiber

**Philippe Chanclou**
Ecole Nationale Supérieure des Télécommunications de Bretagne, Departement d'Optique,
BP226, 22303 Lannion, France

**Monique Thual and Jean lostec**
France Telecom R&D, DTD/PIH, 2 avenue Pierre Marzin, 22307 Lannion, France

A micro-optics has been developed to limit backreflection at the end face of single mode fibers. The measured return-loss at a wavelength of 1.55 µm is as high as 28 dB. This device allows preserving the geometric aspect of a normally cut end face fiber.

**OCIS codes**: (060.2340) Fiber optics components; (110.2760) Gradient-index lenses; (120.5700) Reflection; (120.3620) Lens design.

In transmission line in optical communication, when light is coupled into or out of an optical fiber through a flat cleaved or polished end face of the fiber, reflection may set up disturbing optical systems and devices. Consequently, to obtain the optimum performances of these systems and devices, it is effective to reduce the level of reflection. The primary cause of backreflection or return loss is a discontinuity in the index of refraction of the fiber. The typical Fresnel reflection of flat end face fiber is $3.3 \cdot 10^{-2}$ (-14.7 dB).
To obtain lower optical backreflection, the end face of the fiber should have [1] either:
- (1) anti-reflecting (AR) coating of the fiber end face. Dielectric AR multilayers can be deposited onto the end face of fiber. A power reflectivity of $10^{-3}$ can be achieved.
- (2) lens shaping [2]. The end fiber is heated to form a hemispherical shaping. The level of reflection of this termination has been measured as $10^{-4}$.
- (3) immersion cell. The immersed end face of the fiber in an index matching has a power of reflectivity of $10^{-5}$.
- (4) tilted end face. The end face of the fiber is cleaved or polished with an angle to separate in the angular dimension the reflected light flux from the incident mode. Residual reflections of $10^{-6}$ are obtained.

In this paper, we report a solution to reduce backreflection using micro-optics at the end face of the fiber. The aim of this work is to reduce backreflection level without modifying the geometrical aspect of the fiber without any tilt or bubble lens. We have been looking for a process where the external diameter of 125 µm is maintained along the fiber. This solution using micro-optics may be used singly or in combination with the other methods to achieve high return loss.

This micro-optics consists of an arrangement of graded index (GRIN) fiber and coreless silica fiber of calibrated lengths spliced at the end face of the SMF. The spot size is adjusted by changing the length of GRIN and silica sections. The fabrication process of micro-optics has already been published [3]. First techno-economical analysis shows a very attractive cost of this micro-optics. This is mainly due to the negligible cost of raw materials, the few number of fabrication steps, the collective treatment (by using fiber ribbon), the use of commercially available tools and the high yield of the whole process. We have named this micro-optics GRADISSIMO for GRADed Index, Silica, SIngle-MOde fiber.

The principle of this low backreflection is schematically described in Fig. 1. The incident optical beam from the single mode fiber, is propagated through a free space section (silica) and a graded index medium before to be reflected on the end face. The dependence of the beam parameters can be described by the ray matrix transformation [3, 4]. The reflected beam is back propagated in these two sections before to be coupled in the single mode fiber. The level of backreflection describes only the amplitude fraction reflected back into the incident mode of the fiber. The coupling efficiency of the reflected beam with the fundamental mode $LP_{01}$ of the fiber is function of the length of graded index and silica. This dependence with these sections has been calculated and measured. The level of reflection is a periodical function of the length of GRIN (Pitch/4 = 365 µm).

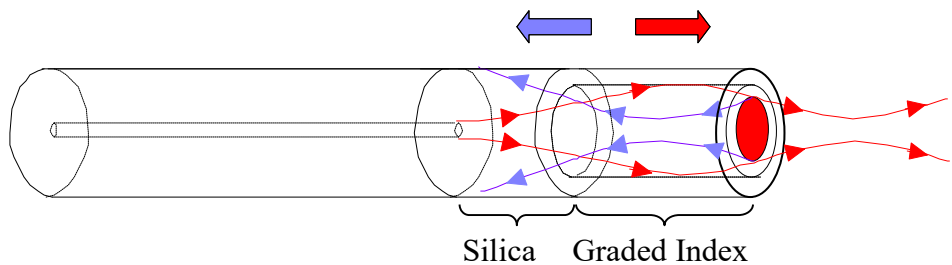

Fig. 1. Schematic illustration of the high return loss scheme that utilizes the micro-optics.

To measure the dependence of the section lengths of the micro-optics on the level of back reflection, we have used a precision White-light INterferometry Reflectometer (WIN-R). This equipment measures the level and location of micro-reflections inside the fiber with a spatial resolution of 50 µm and a return-loss range of 0 to 100 dB. The measurements of the return loss have an accuracy ±2 dB. WIN-R incorporates a fiber source at a wavelength 1550 nm.

As an exemple, Fig. 2 shows a result of backreflection measure. Micro-optics characteristics are 400 µm of silica and 248 µm of GRIN. We measure backreflection level of –30.2 dB at the end face of the micro-optics. The two splicing interfaces of the SMF / silica and silica / GRIN are measured below – 80 dB of reflection (Fig. 3).

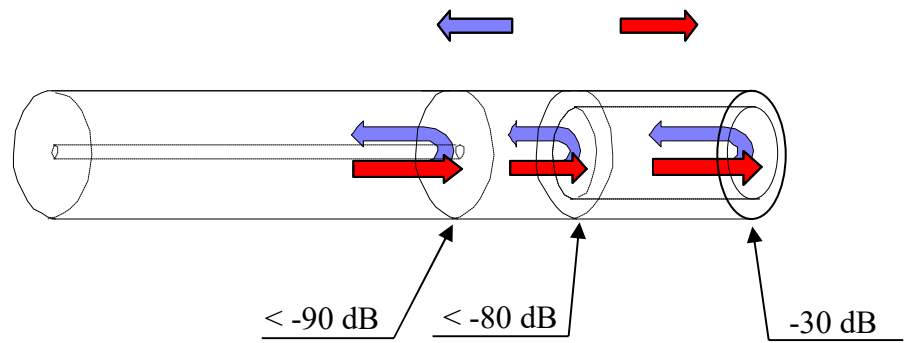

Fig. 3. Schematic view of the measure.

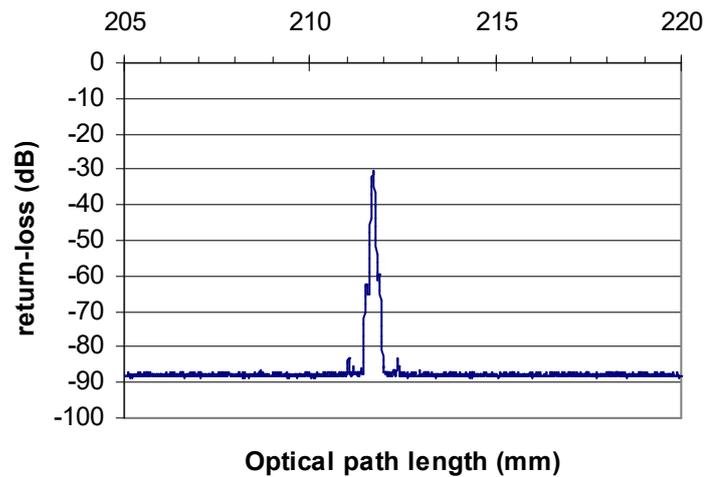

Fig. 2. Experimental measure of backreflection at the end face of the micro-optics.

In the following, we outline measurements of micro-optics reflection. The results are reported in Fig. 4 (a) and (b) as function of GRIN and silica lengths. We show that the contribution of the GRIN length to the reflection is a periodical function. The reflected power from a SMF with a micro-optics without silica ($L_{silica} = 0$ µm) is measured to be below a level line of the forward power oscillating between –14.7 dB and –18 dB (Fig. 4 (a)). For a length of silica of 400 µm, the level oscillates between –14.7 dB and –28 dB (Fig; 4 (b)).
The theoretical curves (solid lines) are compared with the measured values. These curves are in good agreement with the upper limit of measured points. Since no polishing is applied in this work, the end face imperfection should modify the value of the return loss measurements. The larger the optical beam at the end face is, the more important are the influences of these end face imperfections on the level of reflection. So the dispersion of the results is all the more important as the spot size diameter at the end face is biggest. The spot size diameter of the micro-optics versus varying lengths of GRIN and silica sections has already been published [4].

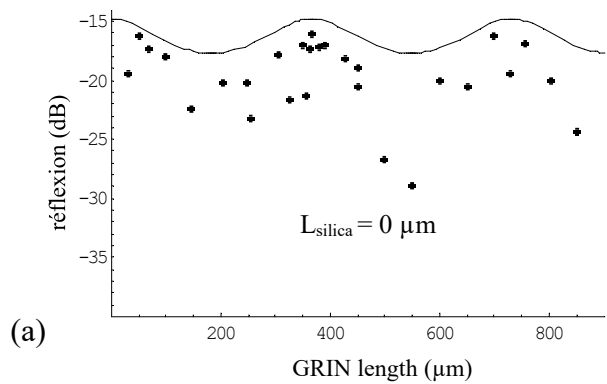

(a)

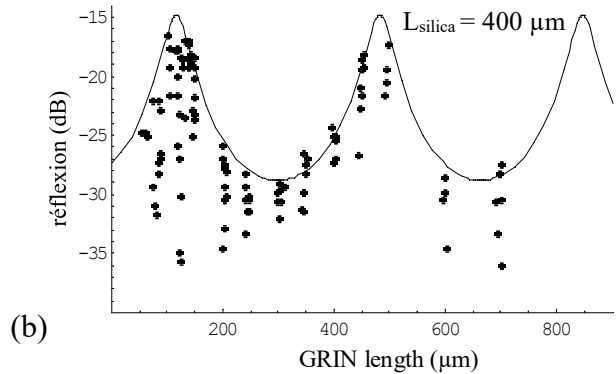

Fig. 4. Backreflection measures as a function of GRIN length for two lengths of silica $L_s = 0$ µm (a) and 400 µm (b). Solid lines: theoretical curves.

In this paper, we present a micro-optics at fiber end to reduce backreflection. The level of reflection can be reduced under –28 dB. The theory is in a good agreement with experimental results to define the upper limit of reflections. The fabrication process needs few simple and collective steps and is particularly suitable for mass production and low cost.
The potential of this device to obtain very high return loss in combination with anti-reflecting coating is expected in the future. This solution allows preserving the geometric aspect of the fiber (no tilt) and simplifies the packaging process.